# Analysis of the homogeneity and heterogeneity of cryptocurrency

Liu Xiao Fan1,2,* Lin Zeng-Zian2 Han Xiaopu3

1 Department of Media and Communication, City University of Hong Kong

2 School of Computer Science and Engineering, Southeast University

3 Alibaba Business School, Hangzhou Normal University

* xf.liu@cityu.edu.hk

## Abstract

Since the advent of Bitcoin in 2008, there have been thousands of cryptocurrencies circulating on exchanges in the world, with a total market value of more than two trillion yuan in 2019. This research analyzes and compares the market price, market transaction volume, on-chain transaction volume, mining difficulty, and public opinion heat of 3607 cryptocurrencies, and aims to reveal the heterogeneity between cryptocurrencies, that is, the development of Balance, and homogeneity (homogeneity), that is, prices are affected by the same market factors, and explore the reasons behind these phenomena.

## 1. Introduction

The cryptocurrency represented by Bitcoin is a new type of transaction object. In the absence of precise economic theory for pricing cryptocurrency, the total market value of thousands of cryptocurrencies can reach a scale of nearly two trillion yuan, and the daily transaction volume can reach hundreds of billion yuan [1]. The financial phenomena embodied in the cryptocurrency market and the driving reasons behind these phenomena have aroused widespread interest among scholars in recent years. This article mainly studies the similarities and differences between thousands of cryptocurrencies and attempts to analyze the reasons for these similarities and differences.

Research on the similarities and differences between cryptocurrencies has mainly appeared in the last five years. The literature covers physics and economics journals and conferences. Here we can only do a simple combing. In 2014, Gandal and Halaburda observed the competition between different encrypted electronic currencies and exchanges, analyzed the influence of network effects on electronic currency market competition, and found that the "winner takes all" effect makes Bitcoin dominate [2]. In 2018, Ciaian et al. found that there was a short-term correlation between the price trends of Bitcoin and other electronic currencies [3]. Yi et al. [4] and Song et al. [5] analyzed the prices of multiple electronic currencies and found that Bitcoin and Ethereum are the foremost market leaders.

The reasons for the fluctuation of electronic currency prices are also an important research topic. In addition to internal market factors, external factors have a significant impact on transaction volume, volatility, and price movements. In 2013, some scholars began to study the relationship between the popularity of online search and the popularity of encyclopedia entry editing and Bitcoin price trends [6]. Since then, there have been scholars linking Bitcoin's trading volume, price trends and online discussions, breaking news events [7, 8], and even risk aversion in gold, securities markets, and other investment products [9, 10] ].

In summary, we can find that scholars have begun to pay attention to the linkages of the cryptocurrency market and the source of the volatility of certain specific currencies. However, they have not yet seen the relationship between the market indicators of all virtual currencies and external influence factors. Because of this, this research aims to analyze the homogeneity and heterogeneity of all cryptocurrencies from a more macroscopic and comprehensive perspective, especially to analyze the source of price heterogeneity in combination with multiple external factors, to solve the above problems.

## 2. The homogeneity of cryptocurrency

There are three primary sources of value of cryptocurrency. One is payment currency represented by Bitcoin, which can only be used as a commodity currency for payment; the other is legal currency anchored currency represented by Tether, whose value is anchored to a specific legal currency. The other is an asset-mapping token represented by Binance Coin, which most of the time represents the equity assets of the project. Except for fiat currency anchored currencies, the value of other cryptocurrencies in the secondary market is challenging to measure. However, through correlation analysis, we can understand whether the price fluctuations of these cryptocurrencies are affected by the same economic factors, to provide clues for further analysis of their value sources.

We captured the price trends of 848 electronic currencies that were actively traded on the CoinMarketCap website from January 1 to November 25, 2018, and converted the price time series of each cryptocurrency i into a logarithmic return time sequence

$$r_i = \log\left(\frac{\text{close}_t}{\text{close}_{t-1}}\right), t \in [2, N]$$

Among them, N is the total number of days in the time range included in the data, and {\rm close}_t is the closing price on day t (the latest price of the day in UTC marked on Coinmarketcap). We calculated the Pearson correlation coefficients of virtual currency i and virtual currency j respectively

$$\rho_{i,j} = \frac{\mathrm{Cov}(r_i, r_j)}{\sigma_{r_i}\sigma_{r_j}}$$

And partial correlation coefficient

$$\rho_{i,j(\mathrm{Bitcoin})} = \frac{\rho_{i,j} - \rho_{i,\mathrm{Bitcoin}}\rho_{j,\mathrm{Bitcoin}}}{\sqrt{1 - \rho_{i,\mathrm{Bitcoin}}^2}\sqrt{1 - \rho_{j,\mathrm{Bitcoin}}^2}}$$

which is the correlation between virtual currency i and virtual currency j after removing the price trend of Bitcoin. In this way, we can judge the impact of market factors other than Bitcoin on virtual currency returns.

Figure 1(a) shows the distribution of the linear correlation coefficient \rho_{i,Bitcoin} between Bitcoin and all other electronic currencies. It can be seen that most of the electronic currency income is positively correlated with Bitcoin. Figure 1(b) shows the distribution of correlation coefficients (blue) and partial correlation coefficients (pink) among all electronic currency returns. It can be found that the linear correlation coefficients \rho_{i,j} between cryptocurrency returns are greater than 0, indicating that they are affected by similar market factors. However, after removing the trend of Bitcoin, the median of the \rho_{i,j(Bitcoin)} distribution of cryptocurrency returns is close to 0, and the distribution is relatively concentrated.

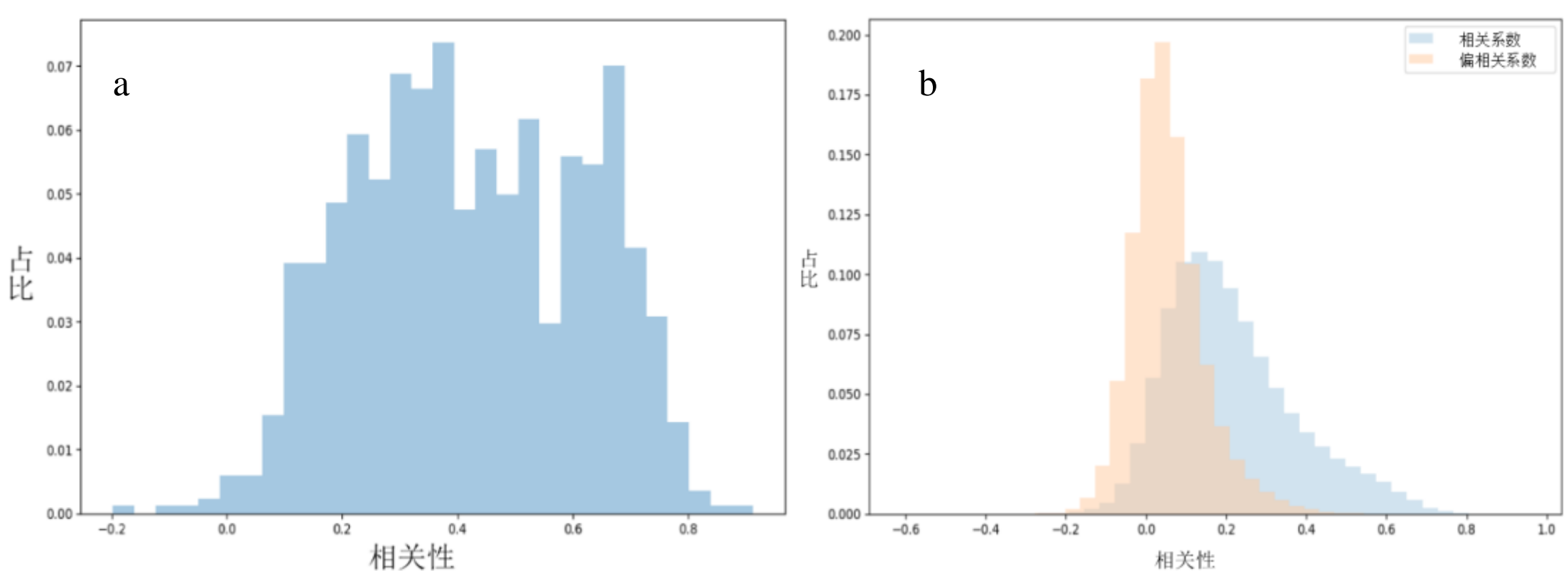

Figure 1: (a) Distribution of linear correlation coefficients between Bitcoin and other virtual currencies, (b) distribution of correlation coefficients and partial correlation coefficients of all electronic currencies

This result shows that the earnings trend of Bitcoin is the anchor of the market, and the overall impact of other market factors on the return rates of other electronic currencies is close to random. In other words, the market's value measurement of cryptocurrency is extremely homogeneous, and all factors that can cause bitcoin price fluctuations will have the same manifestation on all electronic currencies at the same time.

## 3. The heterogeneity of cryptocurrencies

Although the volatility of the cryptocurrency is homogeneous, due to the difference in market recognition, there are apparent differences in the market value, currency price, and trading volume indicators of different cryptocurrencies. We crawled from the CoinGecko website the market value, currency price, 24-hour trading volume, social popularity, and other market indicators of the 3607 cryptocurrencies that were traded on December 16, 2018.

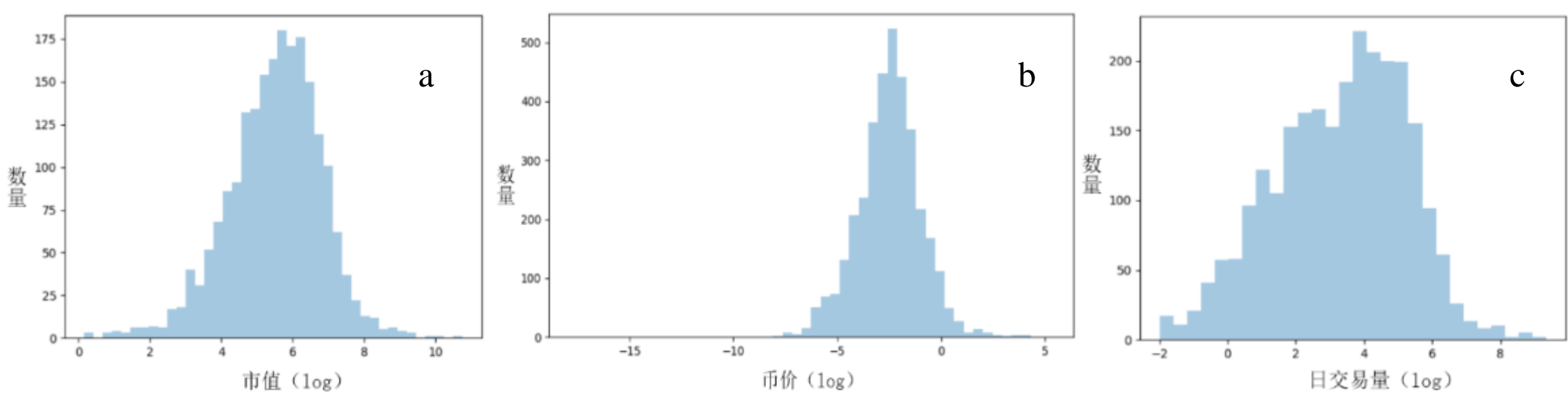

Figure 2: (a) Market value distribution, (b) currency price distribution, (c) 24-hour trading volume distribution of cryptocurrency, in US dollars

Figure 2 (a), (b), (c) respectively show the logarithmic distribution of the market value of cryptocurrencies, the logarithmic distribution of currency price, and the logarithmic distribution of 24-hour transaction volume. It can be found that the market indicators of cryptocurrencies are all close to lognormal distribution, and there is a certain degree of heterogeneity.

We further use the Herfindahl-Hirschman index and the Gini coefficient to quantify the unevenness of the distribution. The Herfindahl-Hirschman Index (Herfindahl-Hirschman Index) is a comprehensive index used by economics to measure industrial concentration. It is used to measure the dispersion of market shares. The calculation formula is as follows.

$$HHI = \frac{\sum_{i=1}^{N} {x_i}^2}{(\sum_{i=1}^{N} x_i)^2}$$

Where $x_i$ represents the sample, and N is the sample size. When the market is monopolized by one company, HHI=1; if all companies are of the same size, $HHI=\frac{1}{N}$, the larger the N, the closer the HHI to zero. The usual way to use HHI is to multiply its value by 10000 to enlarge it. The US Department of Justice (Department of Justice) uses HHI as an indicator to assess the concentration of a particular industry. It believes that $HHI$ less than 1500 is a competitive market, and $HHI$ is between 1500 and 2500. It is a moderately concentrated market, with more than 2500 highly concentrated markets [11]. Gini coefficient (Gini coefficient) is a commonly used international index to measure the sample gap. The calculation formula is as follows.

$$GINI = \frac{2\sum_{i=1}^{N} ix_i}{N\sum_{i=1}^{N} x_i} - \frac{N+1}{N},$$

Where x_i represents the x index of cryptocurrency i, and N is the sample size. It is generally believed that the distribution of indicators when the Gini coefficient reaches 0.5 or more is extremely disparity [12].

Table 1 summarizes the Herfindahl-Hirschman index and Gini coefficient of 3607 cryptocurrencies at the end of 2018 in terms of market value, currency price, and 24-hour trading volume. It can be found that the market value and currency price of electronic money are unevenly distributed, the Gini coefficient is close to 1, and the HHI is also higher than the high concentration standard of 2500; the Gini coefficient of the 24-hour trading volume has also reached 0.99. It can be seen that the cryptocurrency market is already heterogeneous and oligopolistic.

Table 1: HHI and Gini coefficients of 3,607 cryptocurrencies at the end of 2018

表 1：3607 种虚拟电子货币 2018 年末三种市场指标的 *HHI* 和基尼系数

| 指标 | ***HHI*** | 基尼系数 |
|---|---|---|
| 市值 | **3184** | **0.988** |
| 币价 | **4991** | **0.998** |
| **24** 小时市场交易量 | 1194 | **0.990** |

### 4. Analysis of possible reasons for the heterogeneity of cryptocurrency

What causes the heterogeneity of market value, currency price, and trading volume of cryptocurrency in the market? Which user's behavior affects the market value of a particular electronic currency? It is generally believed that there are four distinct camps in the user ecology of cryptocurrency. The first is the technology-led developer camp. They pay attention to the development of blockchain technology and carry out technical discussions around various projects. The second camp is speculators who care about currency price fluctuations in the short term and use various means to arbitrage. The third camps are liberals, that is, early Bitcoin acceptors. The fourth camp is the real users of cryptocurrency. They may be groups of miners or criminals buying illegal products[13]. For different user groups, we collected the number of Reddit subscriptions (representing users who follow technology), Facebook likes, and Twitter followers of the 3607 cryptocurrencies mentioned in the previous section on December 16, 2018 ( On behalf of users who are concerned about speculation). As well as user-related records of the number of transactions on the chain of certain cryptocurrencies that have a self-generated blockchain, and the difficulty of mining.

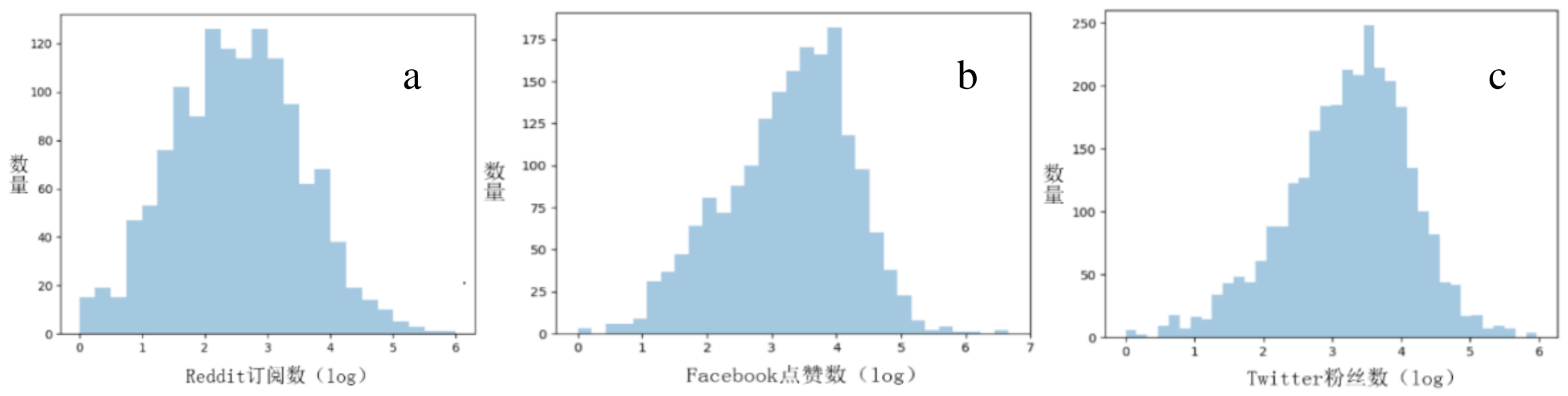

Figure 3: Distribution of social popularity indicators of cryptocurrency, (a) Reddit subscriptions, (b) Facebook likes, (c) Twitter followers

Figure 3 (a), (b), and (c) respectively show the logarithmic distribution of Reddit subscriptions, Facebook likes, and Twitter followers of 3607 cryptocurrencies. The number of Facebook likes, and the number of Twitter followers show a significant right-skewing under the logarithmic distribution, while the number of Reddit subscriptions is closer to the normal distribution under the logarithmic scale. Table 2 shows that although the Gini coefficient is higher than 0.8 in terms of social heat, HHI exceeds the general economic standard.

Table 2: HHI and Gini coefficient of three social popularity indicators for more than 3,000 cryptocurrencies at the end of 2018

| 指标 | HHI | GINI |
|---|---|---|
| Reddit 订阅数 | 369 | **0.804** |

| Facebook 点赞数 | 443 | **0.854** |
|---|---|---|
| Twitter 粉丝数 | 59 | **0.804** |

We further crawled the on-chain data of 257 cryptocurrencies with native blockchains from the blockchain browser website chainz.cryptoid.info and obtained indicators such as the difficulty of mining and the number of transactions on the chain in 24 hours. Table 3 shows the uneven distribution of these two indicators, and they also show apparent heterogeneity.

Table 3: Uneven data distribution indicators of 257 kinds of cryptocurrencies with protogenesis links

| 指标 | HHI | GINI |
|---|---|---|
| 24 小时链上交易数 | 791 | **0.789** |
| 挖矿难度 | **5208** | **0.992** |

In order to explore the possible causes of heterogeneity, we analyze the correlation between on-chain indicators and social popularity indicators and market indicators such as the market value and currency price of cryptocurrencies. As shown in Table 4, the technical discussion heat (number of Reddit subscriptions) of 3607 currencies shows a strong linear correlation with their currency price and market value. The speculative user attention of these currencies (number of likes on Facebook and Twitter followers) number) is not strongly correlated with their market performance. Among the 257 online blockchains, the two indicators directly related to blockchain users, the number of transactions in 24 hours, and the difficulty of mining do not show strong correlations with market indicators. Therefore, we tend to draw the following conclusion: the heterogeneity of the market value of electronic currency is caused by their technological enthusiasm (maturity). The better a blockchain technology, e.g., the more developers, is, the more valuable it is. Large, and the higher the transaction volume. The number of miners or usage has nothing to do with market value.

Table 4: Correlation analysis of cryptocurrency market performance, social popularity, and on-chain transaction indicators

| | 币价 | 市值 |
|---|---|---|
| 24 小时市场交易量 | 0.08 | **0.85** |
| 24 小时链上交易数 | -0.01 | 0.06 |
| 挖矿难度 | -0.00 | -0.01 |
| Reddit 订阅数 | **0.77** | **0.90** |
| Facebook 点赞数 | 0.01 | -0.04 |
| Twitter 粉丝数 | 0.01 | 0.18 |

## 5. Conclusion

Cryptocurrency is a new type of financial product that has formed a substantial global trading market within a short period after its birth. This article has conducted a preliminary analysis of the market performance of more than 3000 cryptocurrencies in this market, such as the correlation between market value, price, and transaction volume, and the conclusion is that they are both homogeneous and heterogeneous. Homogeneity is manifested in the market. Financial factors that can affect the income of cryptocurrencies will be reflected in all cryptocurrencies at the same time and produce the same effect. The heterogeneity is manifested in the extremely uneven distribution of these electronic currency market indicators, such as market value, price, and transaction volume. We tested the relationship between the activity of different types of cryptocurrency users, such as developers, speculators, and offline users on different channels, and the market value of cryptocurrencies, and finally found that the technical discussion is hot, that is, the subscription on Reddit The number has a strong correlation with the market value and currency price. The research in this article discusses the similarities and differences between all cryptocurrencies from a macro perspective and proposes possible reasons to explain these relationships, hoping to inspire follow-up research.